\documentclass[12pt,twocolumn]{iopart}

\expandafter\let\csname equation*\endcsname\relax
\usepackage{multicol}
\expandafter\let\csname endequation*\endcsname\relax

\usepackage{amsmath}
\usepackage{amsfonts}
\usepackage{iopams}
\usepackage{cite}
\usepackage[colorlinks=true, citecolor=blue]{hyperref}
\usepackage{graphicx}
\usepackage{hyperref}
\usepackage{graphics}
\usepackage{amssymb}
\usepackage{color}
\usepackage{subfigure}
\usepackage{bm}
\usepackage{float}
\usepackage[utf8x]{inputenc}

\begin{document}

\title[Quantum correlations and quantum-memory-assisted entropic ...]{Quantum correlations and quantum-memory-assisted entropic uncertainty relation in a quantum dot system}

\author{Soroush Haseli $^{1}$ \footnote{
Corresponding author}}

\address{
$^{1}$ Faculty of Physics, Urmia University of Technology, Urmia, Iran\\

}

\ead{soroush.haseli@uut.ac.ir}

\vspace{10pt}
\begin{indented}
\item[]
\end{indented}

\begin{abstract}
The uncertainty principle is one of the comprehensive and fundamental concept in quantum theory. This principle states that it is not possible to simultaneously measure two incompatible observatories with high accuracy. Uncertainty principle has been formulated in various form. The most famous type of uncertainty relation is expressed based on the standard deviation of observables. In quantum information theory the uncertainty principle can be formulated using Shannon and von Neumann entropy.  Entropic uncertainty relation in the presence of  quantum memory is one of the most useful entropic uncertainty relations. Due to their importance and scalability, solid state systems have received considerable attention nowadays. In this work we will consider a quantum dot system as a solid state system. We will study the quantum correlation and quantum memory assisted entropic uncertainty in this typ of system. We will show that the temperature in of quantum dot system can  affect the quantum correlation and entropic uncertainty bound. It will be observed that the entropic uncertainty bound decreases with decreasing temperature and quantum correlations decreases with increasing the temperature. 
\end{abstract}

%

%
%
%
%

\section{Introduction}\label{Sec1}
According to the fundamental role of quantum correlations in quantum information theory, this subject has been extensively studied in recent years \cite{1,2,3,4,5,6,7,8}. It makes a variety of applications possible in quantum information theory such as quantum teleportation \cite{6,9}, quantum cryptography \cite{10}, quantum dense coding \cite{11}, quantum computing \cite{12,13,14} and quantum communication \cite{15,16}. In previous decades, it has been thought that the only correlation in quantum information theory is entanglement. Hence, different criteria  were introduced to measure the entanglement, including concurrence \cite{2}, von Neumann entropy, negativity and logarithmic negativity \cite{17,18}, entanglement cost \cite{19,20}, entanglement of formation \cite{2}, squashed entanglement \cite{21}, robustness of entanglement for bipartite entanglement \cite{22}, tangle \cite{23,24}, relative entropy \cite{25}, generalized concurrence \cite{26,27,28,29}, geometric measures \cite{30,31,32}, global entanglement \cite{33,34} and Scott measure\cite{35,36,37,38,39}. It has been shown that quantum entanglement does not cover all aspects of quantum correlations \cite{40,41}. Therefore, it was necessary to introduce a new criterion for quantum correlations. Up to now, various measures have been introduced to quantify the quantum correlation and each of them has specific features. Some of these measures are quantum discord (QD) \cite{42,43,44}, geometric QD \cite{45,46,47,48}, global geometric QD \cite{47} and super QD \cite{49,50,51}.  Among the correlation criteria mentioned above, concurrence and QD are used more than other ones. 
The uncertainty principle is another comprehensive and fundamental concept in quantum theory which was proposed by Heisenberg \cite{58}. The uncertainty principle states that it is not possible to measure two incompatible observatories simultaneously and with high accuracy. The uncertainty relation can be written in different form. The most famous type of uncertainty relation is expressed based on the standard deviation of observables \cite{59,60}. It has been shown that the uncertainty principle can be formulated using the quantities describing entropy \cite{61,62}. Entropic uncertainty relation (EUR) in the presence of an additional particle that is used as a quantum memory is one of the most useful entropic uncertainty relations \cite{63}. It is known as quantum memory assisted entropic uncertainty relation (QMA-EUR). Entropic uncertainrt relations have a wide range of applications such as quantum key distribution \cite{64,65},quantum cryptography \cite{10}, and quantum metrology \cite{66}. In recent years, much work has been done to improve the EUR \cite{67,68,69,70,71,72,73,74,75,76,77}. In some works, the relationship between quantum correlations and the QMA-EUR has also been investigated \cite{78,79,80,81,82,83,84,85,86,87,88,89,90,91,92,93,94,95,96,97,98,99,100,101}.

Due to their importance and scalability, solid state systems have received considerable attention nowadays. The main goal in solid state quantum physics is creating and determining quantum correlations between individual electrons. The main motivation for these studies comes from the fact that the recent experimental processes in the field of quantum information has led to experimental realization of one and two-qubit utilization electron spin qubits in quantum dots \cite{52,53,54,55} and coherent control of spins in diamond \cite{56,57}. Quantum-dot devices present a well-controlled object to study the quantum many-body
physics. 

In this work, we will study the QMA-EUR and its relation with QD and concurrence in an isolated quantum
dot, in terms of different parameters of the quantum dot. The paper is structured as follows: In Sec. \ref{sec2}, we review the quantum correlation measures used in this work. In Sec. \ref{sec3}, the issue of uncertainty principle will be reviewed . In Sec. \ref{sec4}, the quantum dot system will be described. In Sec. \ref{sec5}, we will study the QMA-EUR and quantum correlation in quantum dot system. Finally, conclusions are presented in Sec.\ref{sec6}.
\section{Concurrence and quantum discord}\label{sec2}
As mentioned in the introduction, there are several criteria for measuring quantum correlations. In this work, we use two practical and optimal criteria, namely concurrence and QD, to measure the quantum correlation in quantum dot system. For a bipartite quantum system the concurrence is given by \cite{102}
\begin{equation}
C=\max\lbrace 0, \lambda_1-\lambda_2-\lambda_3-\lambda_4\rbrace,
\end{equation} 
where $\lambda_i$'s are the eigenvalues, in decreasing order, of the Hermitian matrix $R=\sqrt{\sqrt{\rho_{AB}}\tilde{\rho_{AB}}\sqrt{\rho_{AB}}}$ with $\tilde{\rho_{AB}}=(\sigma_y \otimes \sigma_y) \rho_{AB}^{\star} (\sigma_y \otimes \sigma_y)$, where $\rho_{AB}^{\star}$ is the complex conjugate of  $\rho_{AB}$ and $\sigma_y$ is the
$y$-component of the Pauli matrices. If the density matrix of the quantum system has an X-structure form in the computational basis $\{|00\rangle,|01\rangle,|10\rangle,|11\rangle\}$ i.e.
\begin{equation}\label{state}
\rho_{A B}=\left(\begin{array}{cccc}
\rho_{11} & 0 & 0 & \rho_{14} \\
0 & \rho_{22} & \rho_{23} & 0 \\
0 & \rho_{32} & \rho_{33} & 0 \\
\rho_{41} & 0 & 0 & \rho_{44}
\end{array}\right),
\end{equation}
then the concurrence can be obtained as 
\begin{equation}\label{concurrence}
C=2 \max{0,C_1,C_2},
\end{equation}
where $C_1=\left|\rho_{23}\right|-\sqrt{\rho_{11} \rho_{44}}$ and $C_{2}=\left|\rho_{14}\right|-\sqrt{\rho_{22} \rho_{33}}$. For the state given in Eq.(\ref{state}), the QD can be obtained as  \cite{42,43}
\begin{equation}\label{discordeq}
QD=\min(Q_1,Q_2),
\end{equation}
where 
\begin{equation}
Q_{j}=\mathrm{H}\left(\rho_{11}+\rho_{33}\right)+\sum_{i=1}^{4} \eta_{i} \log _{2} \eta_{i}+D_{j},
\end{equation}
in which $\eta$'s are eigenvalues of density matrix and 
\begin{eqnarray}
D_1&=&H(\alpha),\nonumber \\
D_2&=&-\sum_i \rho_{ii} \log_2 \rho_{ii} + H(\rho_{11}+\rho_{33}),
\end{eqnarray}
where $H(x)=-x \log_2 x -(1-x)\log_2 (1-x)$ and $\alpha=(1+\tau)/2$ with
\begin{equation}
\tau=\sqrt{\left[1-2\left(\rho_{33}+\rho_{44}\right)\right]^{2}+4\left(\left|\rho_{14}\right|+\left|\rho_{23}\right|\right)^{2}}.
\end{equation} 
\section{Quantum memory assisted entropic uncertainty relation}\label{sec3}
The principle of uncertainty is one of the fundamental features of quantum theory first introduced by Heisenberg \cite{58}. In Refs. \cite{59,60}, Schrodinger and Robertson have provided the uncertainty relation based on standard deviation of two incompatible observable $Q$ and $R$ as 
\begin{equation}\label{Robertson}
\Delta Q\Delta R\geq \frac{1}{2}|\langle \left[Q, R\right]\rangle |,
\end{equation}
where $\Delta X=\sqrt{\langle X^{2} \rangle- \langle X \rangle^{2}}$ with $X \in \lbrace Q,R \rbrace$ is the the standard deviation of  $X$, $\langle X \rangle $ shows the
expectation value of operator $X$ and $\left[Q, R \right]=QR-RQ $. The lower bound of Eq.(\ref{Robertson}) depends on the state of the system, which is a defect for this uncertainty relation. In order to solve this problem, the uncertainty relation in terms of Shannon entropy was defined as follows \cite{61,62}
\begin{equation}\label{Rob}
H(Q) + H(R) \geq \log_2 \frac{1}{c}
\end{equation}
where $H(Q)=\sum_i p_i \log_2 p_i$ and $H(R)=\sum_j m_j \log_2 m_j$ are the Shannon entropy, $p_i=\langle q_i|\rho|q_i \rangle$, $m_j=\langle r_j|\rho|r_j \rangle$ and $c=\max_{i,j}\lbrace |\langle q_i | r_j \rangle |^{2} \rbrace$ where $|q_i\rangle$ and $|r_j\rangle$ are  eigenstates of observables $Q$ and $R$, respectively. The EUR was expanded by Bertha by considering an additional quantum system as quantum memory. This type of EUR known as QMA-EUR and is given by \cite{63}
\begin{equation}\label{berta}
S(Q|B)+S(R|B)\geq \log_2 \frac{1}{c}+S(A|B),
\end{equation}
where $S(Q|B)=S(\rho^{QB})-S(\rho^{B})$ and $S(R|B)=S(\rho^{RB})-S(\rho^{B})$ are the conditional von-Neumann entropies of the post measurement states
\begin{equation}\begin{array}{l}
\rho^{Q B}=\sum_{i}\left(\left|q_{i}\right\rangle\left\langle q_{i}\right| \otimes \mathbb{I}\right) \rho^{A B}\left(\left|q_{i}\right\rangle\left\langle q_{i}\right| \otimes \mathbb{I}\right), \\
\rho^{R B}=\sum_{j}\left(\left|r_{j}\right\rangle\left\langle r_{j}\right| \otimes \mathbb{I}\right) \rho^{A B}\left(\left|r_{j}\right\rangle\left\langle r_{j}\right| \otimes \mathbb{I}\right),
\end{array}\end{equation}
and $S(A|B)=S(\rho^{AB})-S(\rho^{B})$ is the conditional von
Neumann entropy. In general QMA-EUR can be explained by a two-player game between Alice and Bob. At the beginning of the game, Bob prepares a bipartite and correlated state $\rho_{AB}$ then he sends part $A$ to Alice and keeps another part $B$ by himself. Part $B$ is used as the quantum memory. In the next step Alice and Bob Reach an agreement on measurement of two observables $Q$ and $R$. Alice does her measurement on part $B$ and declares
her choice of the measurement to Bob via classical communication. Bob tracks to minimize his uncertainty about the outcome of
Alice  measurement .

In Ref.\cite{73}, Adabi et al. provided, the new bound for QMA-EUR which is tighter than Bertha's lower bound. Their QMA-EUR is given by
\begin{equation}\label{Adabi}
S(Q | B)+S(R | B) \geq \log _{2} \frac{1}{c}+S(A | B)+\max \{0, \delta\}
\end{equation} 
where 
\begin{equation}
\delta=I(A ; B)-(I(Q ; B)+I(R ; B))
\end{equation}
and 
\begin{equation}\label{Holevo}
I(X;B)=S(\rho^{B})-\sum_x p_x S(\rho_x^{B}), \quad X \in \lbrace Q, R\rbrace,
\end{equation}
Eq.\ref{Holevo} is known as Holevo quantity, $p_x=tr_{AB}(\Pi_{x}^{A}\rho^{AB}\Pi_x^{A})$ is the probability of x-th outcome and $\rho_x^{B}=\frac{tr_{A}(\Pi_{x}^{A}\rho^{AB}\Pi_x^{A})}{p_x}$  is the state of the Bob after the measurement of $X$ by Alice. In this work we will use Adabi's QMA-EUR in our calculations. In this work, we also choose $Q=\sigma_x$ and $R=\sigma_z$ and investigate the relation between quantum correlation and entropic uncertainty bound (EUB). Since the mutual information and corresponding Holevo quantity  are obtained as 
\begin{eqnarray}\label{mu}
I (A;B)&=&-(\rho_{11}+\rho_{22})\log_{2}(\rho_{11}+\rho_{22}) \nonumber \\
&-&(\rho_{33}+\rho_{44})\log_{2}(\rho_{33}+\rho_{44}) \nonumber \\
&-&(\rho_{11}+\rho_{33})\log_{2}(\rho_{11}+\rho_{33}) \nonumber \\
&-&(\rho_{22}+\rho_{44})\log_{2}(\rho_{22}+\rho_{44}) \nonumber \\
&+&\sum_{i} \eta_{i} \log_2 \eta_i, 
\end{eqnarray}
\begin{eqnarray}\label{hol1}
I(Z;B)&=&-(\rho_{11}+\rho_{22})\log_{2}(\rho_{11}+\rho_{22}) \nonumber \\
&-&(\rho_{33}+\rho_{44})\log_{2}(\rho_{33}+\rho_{44}) \nonumber \\
&-&(\rho_{11}+\rho_{33})\log_{2}(\rho_{11}+\rho_{33}) \nonumber \\
&-&(\rho_{22}+\rho_{44})\log_{2}(\rho_{22}+\rho_{44}) \nonumber \\
&+&\sum_i \rho_{ii}\log_2 \rho_{ii},
\end{eqnarray}
\begin{eqnarray}\label{hol2}
I(X;B)&=&1-(\rho_{11}+\rho_{33})\log_{2}(\rho_{11}+\rho_{33}) \nonumber \\
&-&(\rho_{22}+\rho_{44})\log_{2}(\rho_{22}+\rho_{44}) \nonumber \\
&+&\sum_{i} \xi_{i} \log_2 \xi_i,
\end{eqnarray}
 where $\xi_1=\xi_2=(1-k)/4$ and $\xi_3=\xi_4=(1+k)/4$ and $k=\sqrt{4(\rho_{14}+\rho_{23})(\rho_{23}+\rho_{41})+(1-2(\rho_{22}+\rho_{44}))^{2}}$.

\section{Quantum dot system}\label{sec4}
The quantum dot universal Hamiltonian with the magnetic field is given by \cite{Aleiner,Qin,Berrada}
\begin{eqnarray}\label{Hamiltonian}
H&=&\sum_{ns}\epsilon_n d_{ns}^{\dag}d_{ns}-E_s \hat{S}_{tot}^{2}-E_z S^{Z} \nonumber \\
&+& E_c(\hat{N}-N_0),
\end{eqnarray}
where $\hat{N}=\sum_{ns} d_{ns}^{\dag} d_{ns}$  is the total number operator of electrons in the dot  and $N_0$ can be controlled via a nearby gate voltage. Here, we assume that the dot is tuned into a Coulomb-blockade valley with an even integer electron number ($N_0=2$). So, the two active orbital levels is labeled with $n=\pm1$. The level spacing $\delta=\epsilon_{n=+1}-\epsilon_{n=-1}$ between the last filled and first
empty orbital level is tunable. $\delta$ can be controlled using an externally applied magnetic field. $\hat{S}_{tot}=\frac{1}{2}\sum_{nss^{\prime}} d_{ns}^{\dag}\sigma_{ss^{\prime}}d_{ns^{\prime}}$ is the operators of total spin occupying the spins  $s= \vert \uparrow \rangle $ or $\vert \downarrow \rangle $.  $E_s$, $E_c$ and $E_z$ represents the exchange, charging and Zeeman energies respectively.  The Hamiltonian in Eq.(\ref{Hamiltonian}) shows the
electron-electron interaction. This interaction is almost weak at the mean field level.  If the level spacing $\delta$ is close enough the system will form triplet states to obtain energy from the Hund's rule by regularizing the level occupancy. The lowest energy singlet state and the three components
of the competing triplet state can be defined by means of the total spin quantum number $S=0,1$ and its z-projection $S^{Z}$ \cite{Pustilnik}
\begin{equation}\label{h}
\begin{array}{l}
|1,1\rangle=d_{+1 \uparrow}^{+} d_{-1 \uparrow}^{+}|0\rangle, \\
|1,0\rangle=\frac{1}{\sqrt{2}}\left(d_{+1 \uparrow}^{+} d_{-1 \downarrow}^{+}+d_{+1 \downarrow}^{+} d_{-1 \uparrow}^{+}\right)|0\rangle, \\
|1,-1\rangle=d_{+1 \downarrow}^{+} d_{-1 \downarrow}^{+}|0\rangle, \\
|0,0\rangle=d_{-1 \uparrow}^{+} d_{-1 \downarrow}^{+}|0\rangle,
\end{array}\end{equation}
where $\vert 0 \rangle$ is  ground state of the dot with $N_0-2$ electrons. The transition between states in the above equations can be described by the following operator
\begin{equation}\mathbf{S}_{n n^{\prime}}=\frac{1}{2} \mathcal{P} \sum_{s s^{\prime}} d_{n s}^{+} \sigma_{s s^{\prime}} d_{n^{\prime} s^{\prime}} \mathcal{P}\end{equation}
where $\mathcal{P}=\sum_{s, s^{z}}\left|S, S^{z}\right\rangle\left\langle S, S^{z}\right|$ is the projection  onto  the lowenergy multiplet  in Eq.(\ref{h}). Near the singlet–triplet transition, the
two-electron quantum dot acts as a bipartite system.  Considering the existing correspondence, it is possible to define the relations between the states of two fictitious
$1/2$-spins and the states in Eq.(\ref{h}) as\cite{Pustilnik}

\begin{equation}\begin{array}{l}
|1,1\rangle \Leftrightarrow\left|\uparrow_{1} \uparrow_{2}\right\rangle, \\
|1,0\rangle \Leftrightarrow \frac{1}{\sqrt{2}}\left(\left|\uparrow_{1} \downarrow_{2}\right\rangle+\left|\downarrow_{1} \uparrow_{2}\right\rangle\right), \\
|1,-1\rangle \Leftrightarrow\left|\downarrow_{1} \downarrow_{2}\right\rangle, \\
|0,0\rangle \Leftrightarrow \frac{1}{\sqrt{2}}\left(\left|\uparrow_{1} \downarrow_{2}\right\rangle-\left|\downarrow_{1} \uparrow_{2}\right\rangle\right).
\end{array}\end{equation}
Clearly, in terms of these spins the reduced Hamiltonian of isolated dot model given
by Eq.(\ref{Hamiltonian}) can be written as
\begin{equation}
\hat{H}=\frac{k_{0}}{4} \hat{\mathbf{S}}_{1} \cdot \hat{\mathbf{S}}_{2}-\gamma B_{0} \hat{S}^{Z},\end{equation}
where $\hat{\mathbf{S}}_{1,2}$ are the spin operators, $k_0=\delta-2E_s$ is the deference between the energy of the singlet and triplet states. $\gamma$ is the gyromagnetic ratio and $B_0$ is the magnetic field.
In the following, we set $\hbar =1$ and the Boltzmann constant $K=1$ it to simplify the calculations. The eigenstate and eigenvalues of the  reduced Hamiltonian can be obtained as 
\begin{equation}\begin{array}{l}
H\left|\psi_{1}\right\rangle=E_{1}\left|\psi_{1}\right\rangle=\left(\frac{k_{0}}{16}+\gamma B_{0}\right)\left|\downarrow_{1} \downarrow_{2}\right\rangle, \\
H\left|\psi_{2}\right\rangle=E_{2}\left|\psi_{2}\right\rangle=\left(\frac{k_{0}}{16}-\gamma B_{0}\right)\left|\uparrow_{1} \uparrow_{2}\right\rangle, \\
H\left|\psi_{3}\right\rangle=E_{3}\left|\psi_{3}\right\rangle=\frac{k_{0}}{16} \frac{1}{\sqrt{2}}\left(\left|\uparrow_{1} \downarrow_{2}\right\rangle+\left|\downarrow_{1} \uparrow_{2}\right\rangle\right), \\
H\left|\psi_{4}\right\rangle=E_{4}\left|\psi_{4}\right\rangle=\frac{-3 k_{0}}{16} \frac{1}{\sqrt{2}}\left(\left|\uparrow_{1} \downarrow_{2}\right\rangle-\left|\downarrow_{1} \uparrow_{2}\right\rangle\right),
\end{array}\end{equation}
where $\vert \psi_3 \rangle $ and $\vert\psi_4 \rangle$ are maximally entangled Bell states, while $\vert \psi_1 \rangle$ and $\vert \psi_2 \rangle$ are product state with zero entanglement. 
\section{Quantum correlation and entropic uncertainty relation in quantum dot system}\label{sec5}
The density matrix describing the quantum dot at thermal
equilibrium is defined as a statistical mixture mixture of Hamiltonian eigenstates
\begin{equation}
\rho_T=\frac{1}{\mathcal{Z}}\sum_i \exp(-\frac{E_i}{T}) \vert \psi_i \rangle \langle \psi_i \vert,
\end{equation}
where $\mathcal{Z}=tr(\exp(-H/T))$ is the partition function. In the standard basis $\{|00\rangle,|01\rangle,|10\rangle,|11\rangle\}$ the density matrix $\rho_T$ has X-structure and it can be written as 
\begin{equation}\rho_{T}=\frac{1}{\mathcal{Z}}\left(\begin{array}{llll} 
u & 0 & 0 & 0 \\
0 & w & y & 0 \\
0 & y & w & 0 \\
0 & 0 & 0 & v
\end{array}\right),
\end{equation}
where the matrix elements are given by
\begin{equation}
\begin{aligned}
u &=\exp \left(-\frac{k_{0}-16 \gamma B_{0}}{16 T}\right), \\
w &=\frac{1}{2}\left[\exp \left(-\frac{k_{0}}{16 T}\right)+\exp \left(\frac{3 k_{0}}{16 T}\right)\right], \\
y &=\frac{1}{2}\left[\exp \left(-\frac{k_{0}}{16 T}\right)-\exp \left(\frac{3 k_{0}}{16 T}\right)\right], \\
v &=\exp \left(-\frac{k_{0}+16 \gamma B_{0}}{16 T}\right),
\end{aligned}
\end{equation}
and the partition function is $\mathcal{Z}=u+v+2w$. From Eq.(\ref{concurrence}), the concurrence of the quantum dot system is obtained as 
\begin{equation}
C(\rho_T)=\frac{2}{Z}\max \lbrace |y| - \sqrt{uv} \rbrace.
\end{equation}
From Eq.(\ref{discordeq}), the QD of the Quantum dot system is obtained as 
\begin{equation}
QD=\min \lbrace Q_1, Q_2\rbrace
\end{equation}
with 
\begin{eqnarray}
Q_j=H(u+w)-S(\rho_T)+D_j
\end{eqnarray}
where $S(\rho_T)$ is the von Neumann entropy of the quantum dot system and
\begin{eqnarray}
D_1&=&H(\frac{1+\sqrt{(1-2(w+v))^{2}+4|y|^{2}}}{2}),  \\
D_2&=&-u \log_2 u -2w \log_2 w -v \log_2 v + H(u+w). \nonumber
\end{eqnarray}
In a similar way, by  substituting the elements of the density matrix $\rho_T$ in Eqs.(\ref{mu}), (\ref{hol1}) and (\ref{hol2}), the lower limit of the uncertainty relationship can be obtained.
\begin{figure}[H]
  \centering
  \includegraphics[width=0.31\textwidth]{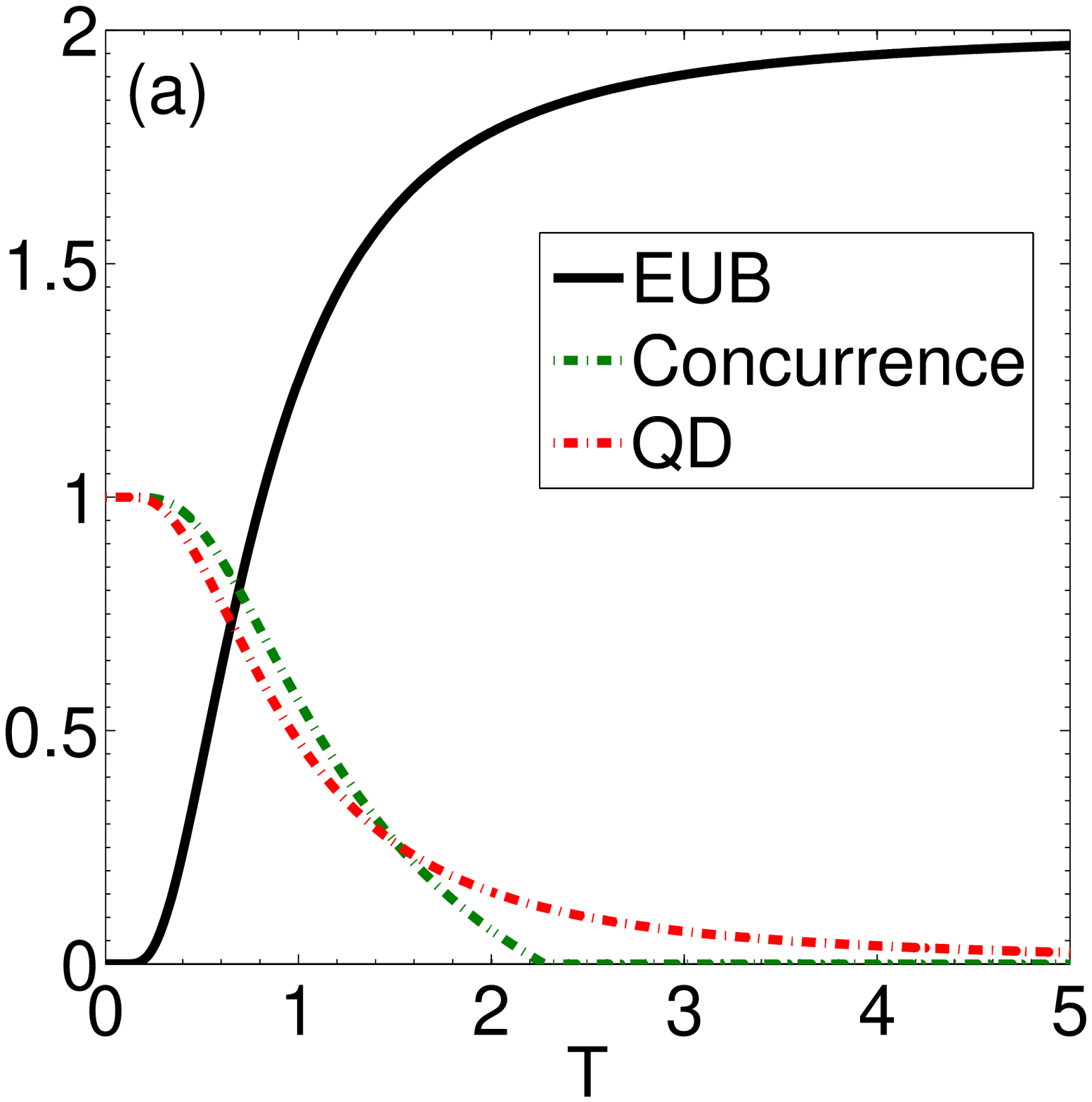}\quad
  \includegraphics[width=0.31\textwidth]{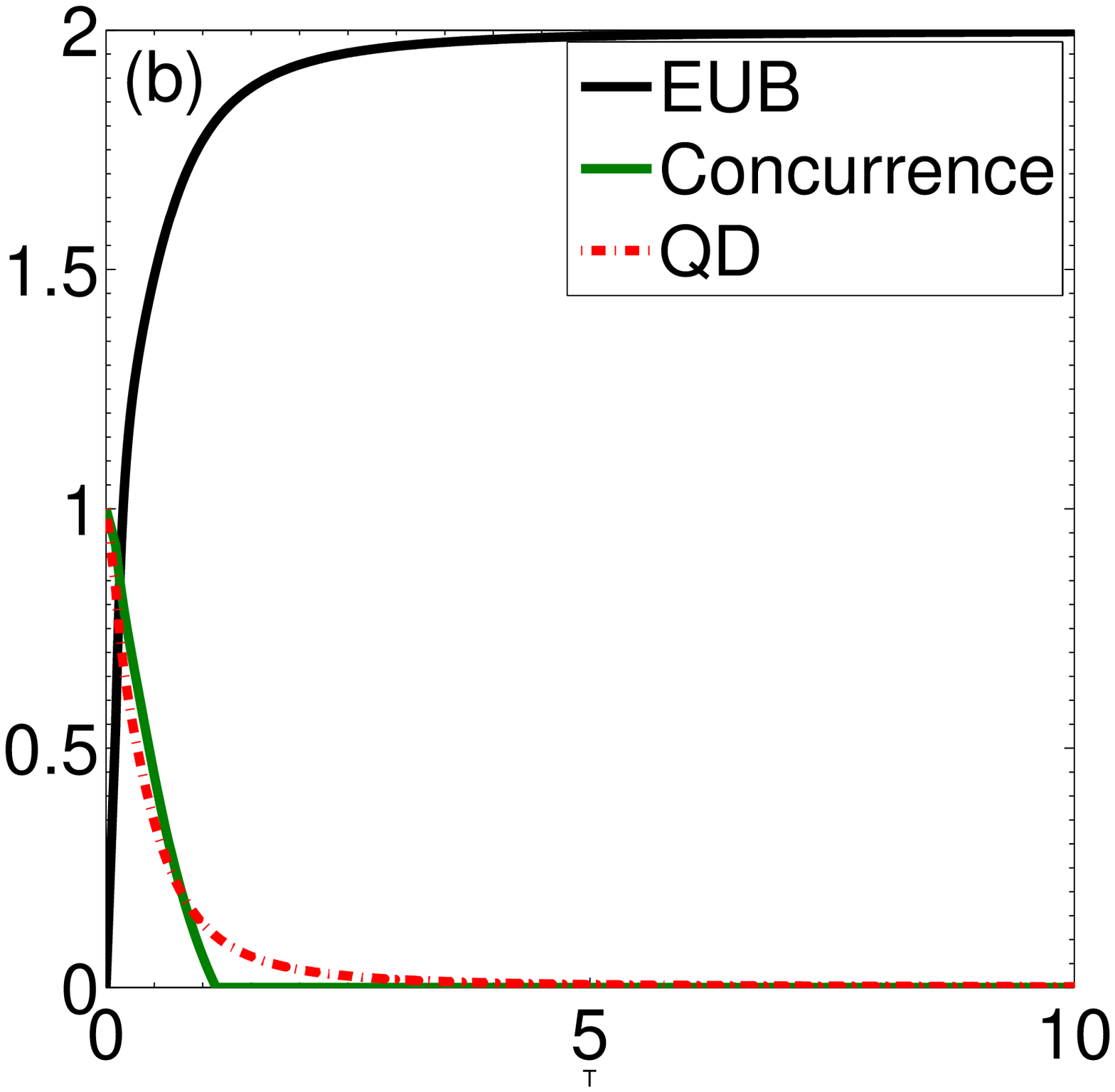}\quad
  \includegraphics[width=0.31\textwidth]{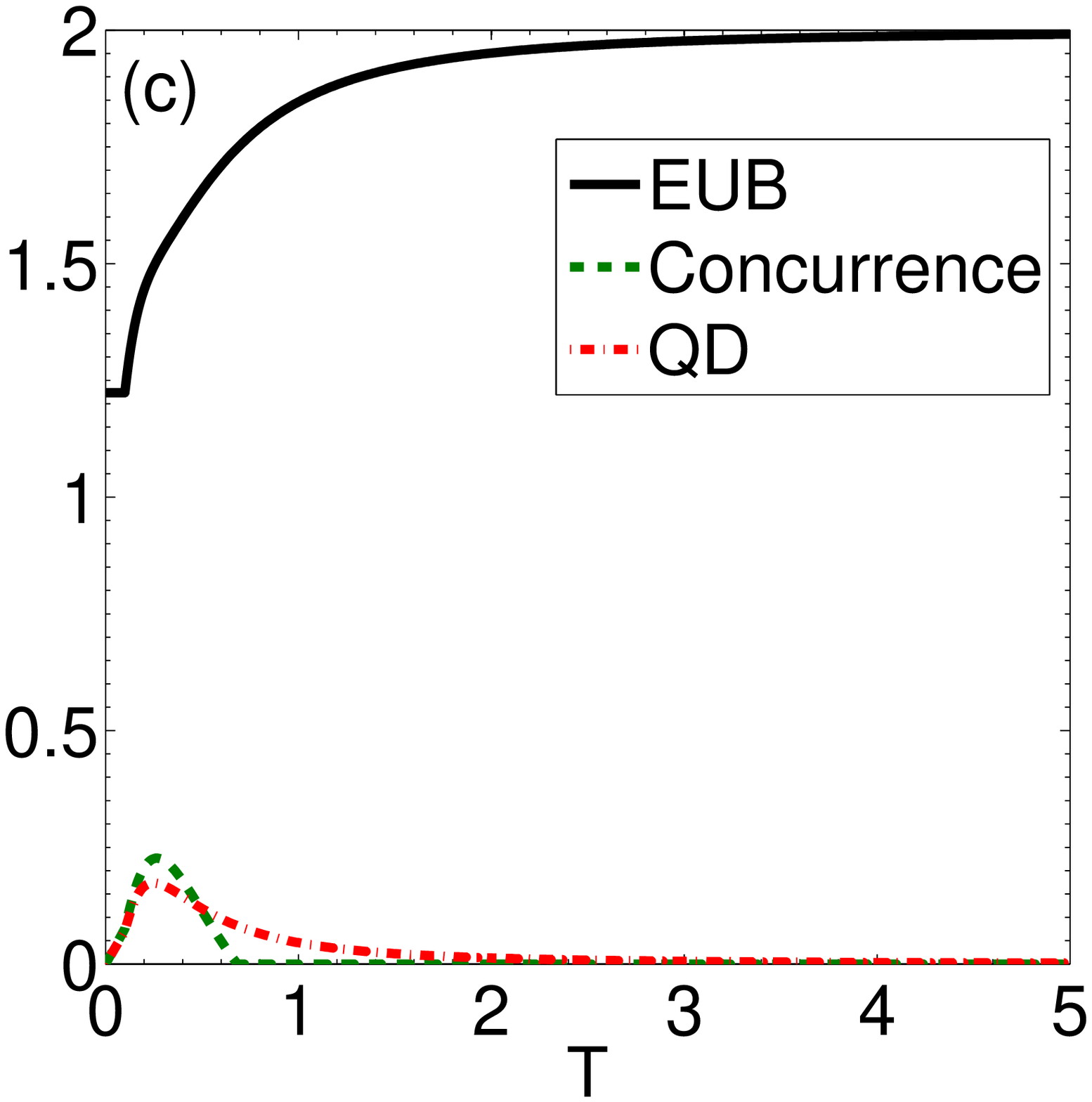}
\caption{Entropic uncertainty bound, concurrence and quantum discord for an isolated quantum
dot in terms of temperature with $B_0=1$, $\gamma=1$ (a)$k_0 =10$. (b) $k_0=5$. (c) $k_0=3$.}
\label{figure1}
\end{figure}
\begin{figure}[H]
  \centering
  \includegraphics[width=0.31\textwidth]{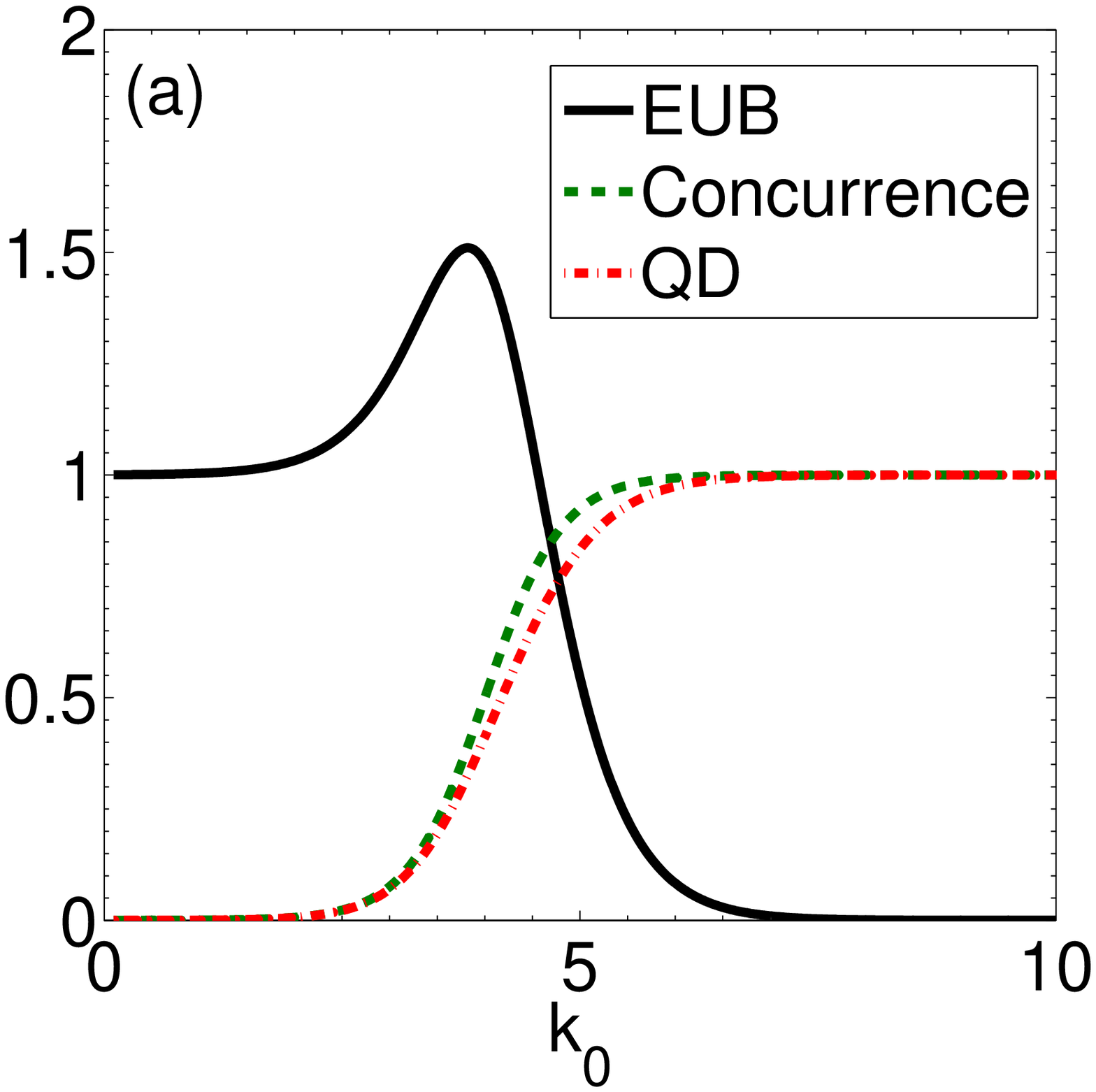}\quad
  \includegraphics[width=0.31\textwidth]{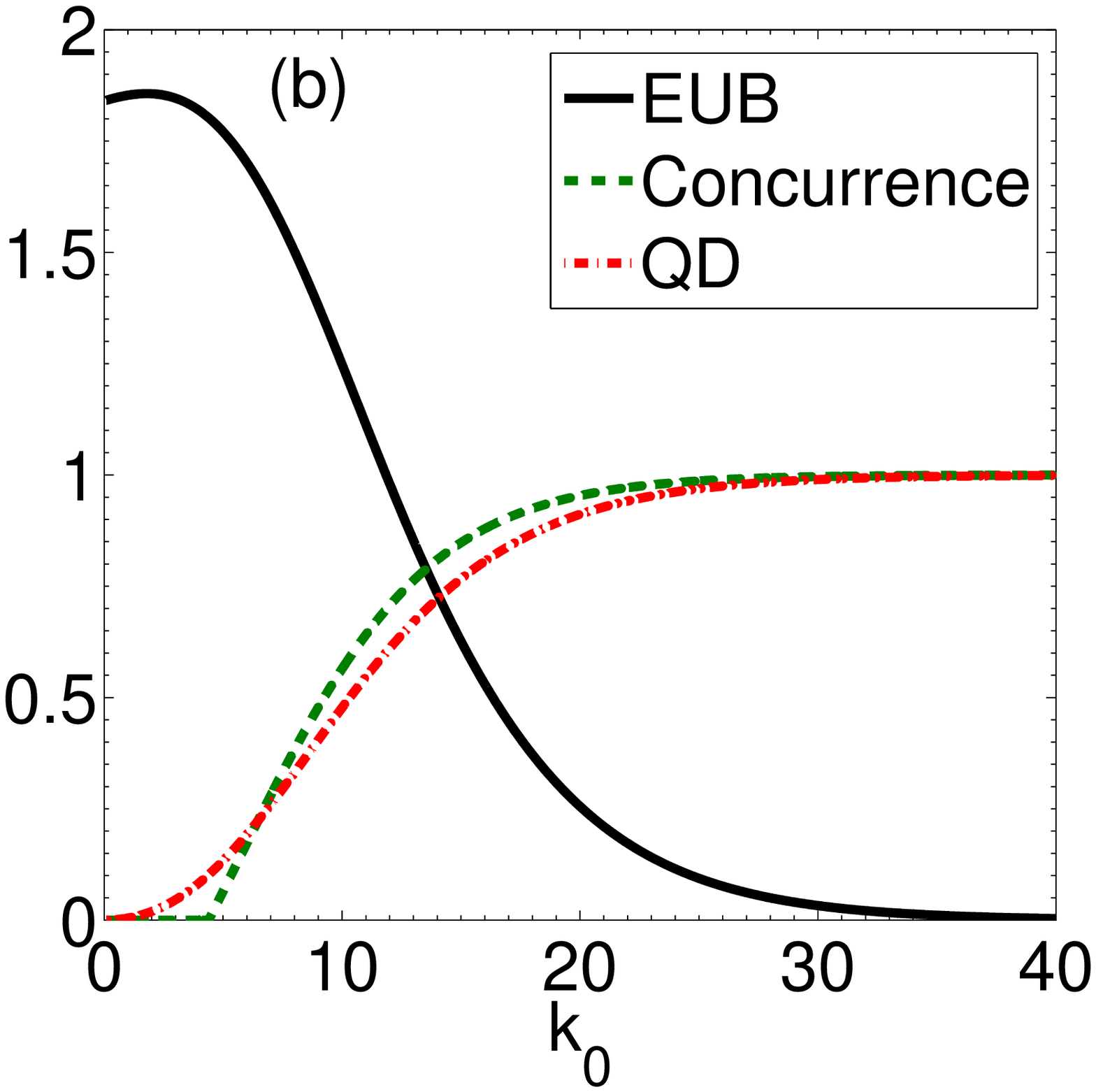}\quad
  \includegraphics[width=0.31\textwidth]{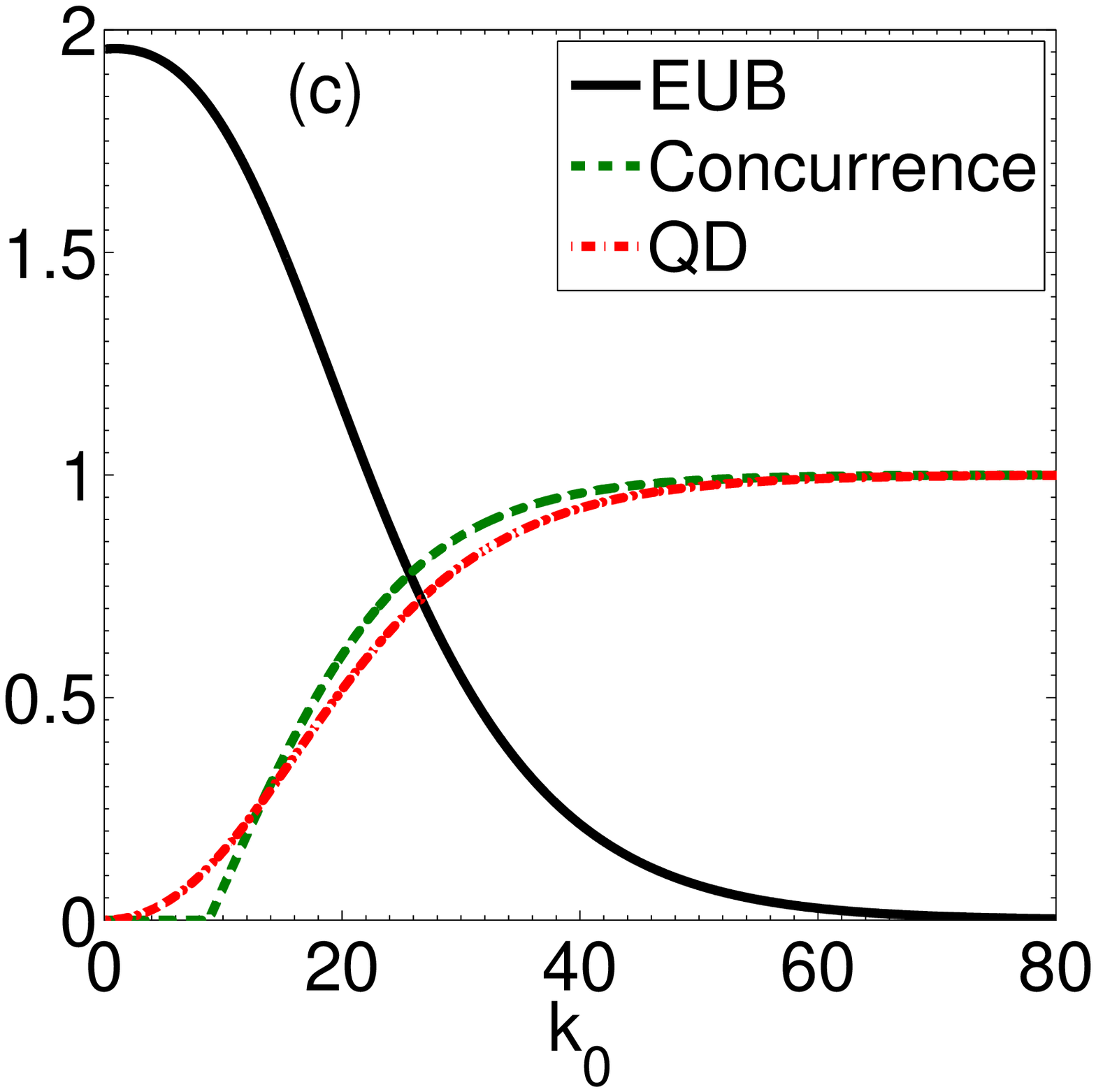}
\caption{Entropic uncertainty bound, concurrence and quantum discord for an isolated quantum
dot in terms of the deference between the energy of the singlet and triplet states $k_0$ with $B_0=1$, $\gamma=1$ (a)$T =0$. (b) $T=1$. (c) $T=2$.}
\label{figure2}
\end{figure}
\begin{figure}[H]
  \centering
  \includegraphics[width=0.31\textwidth]{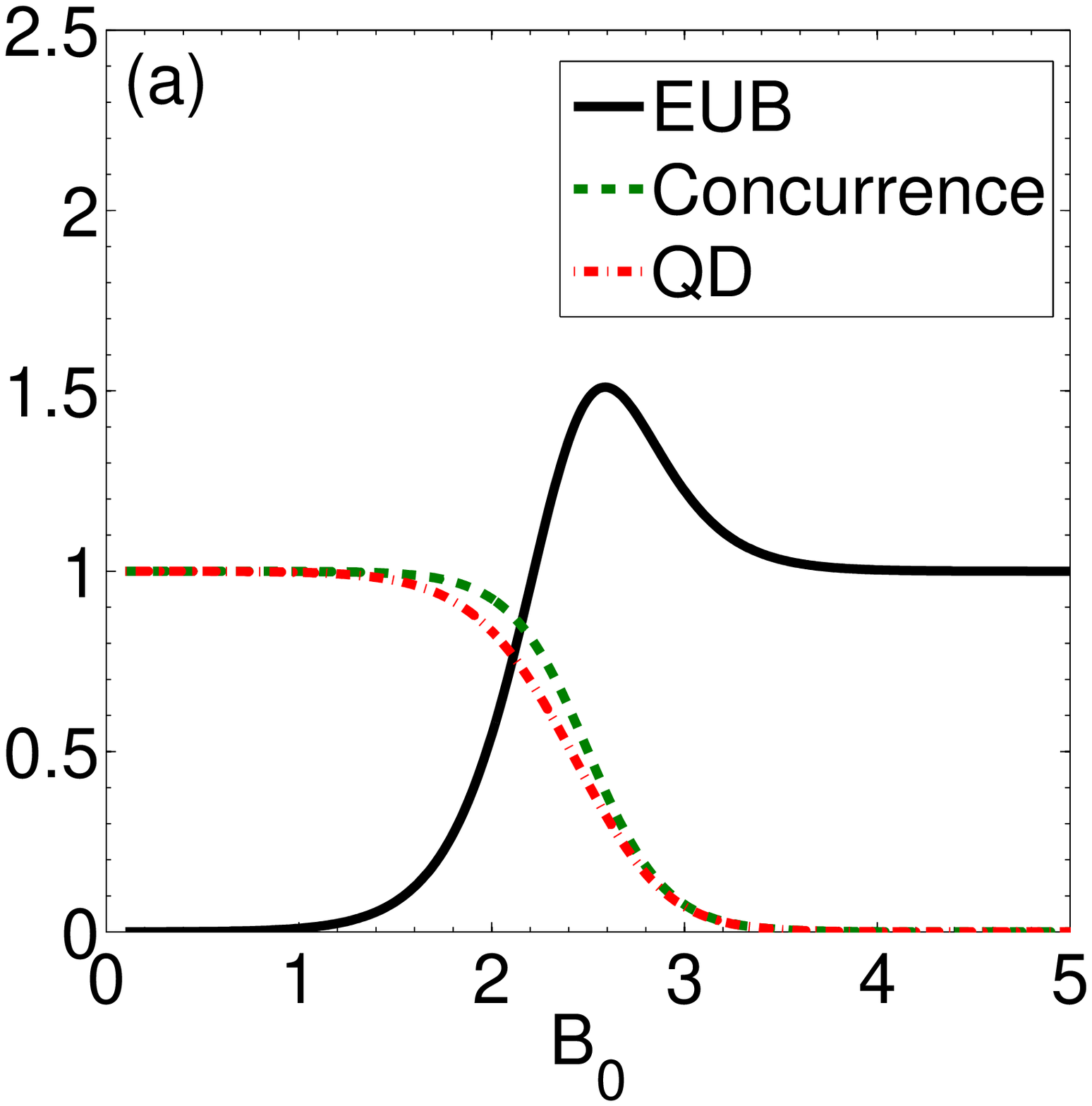}\quad
  \includegraphics[width=0.31\textwidth]{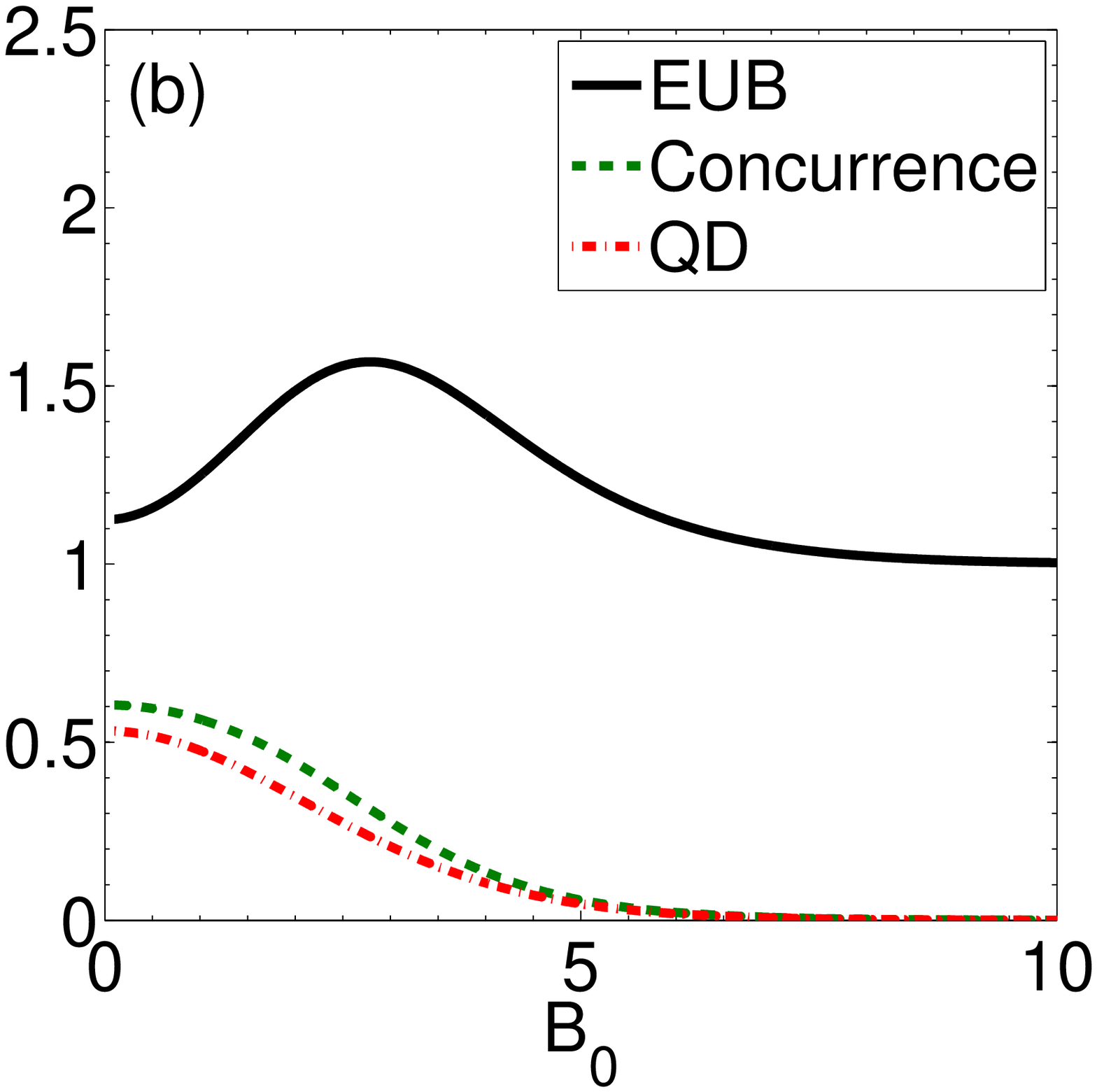}\quad
  \includegraphics[width=0.31\textwidth]{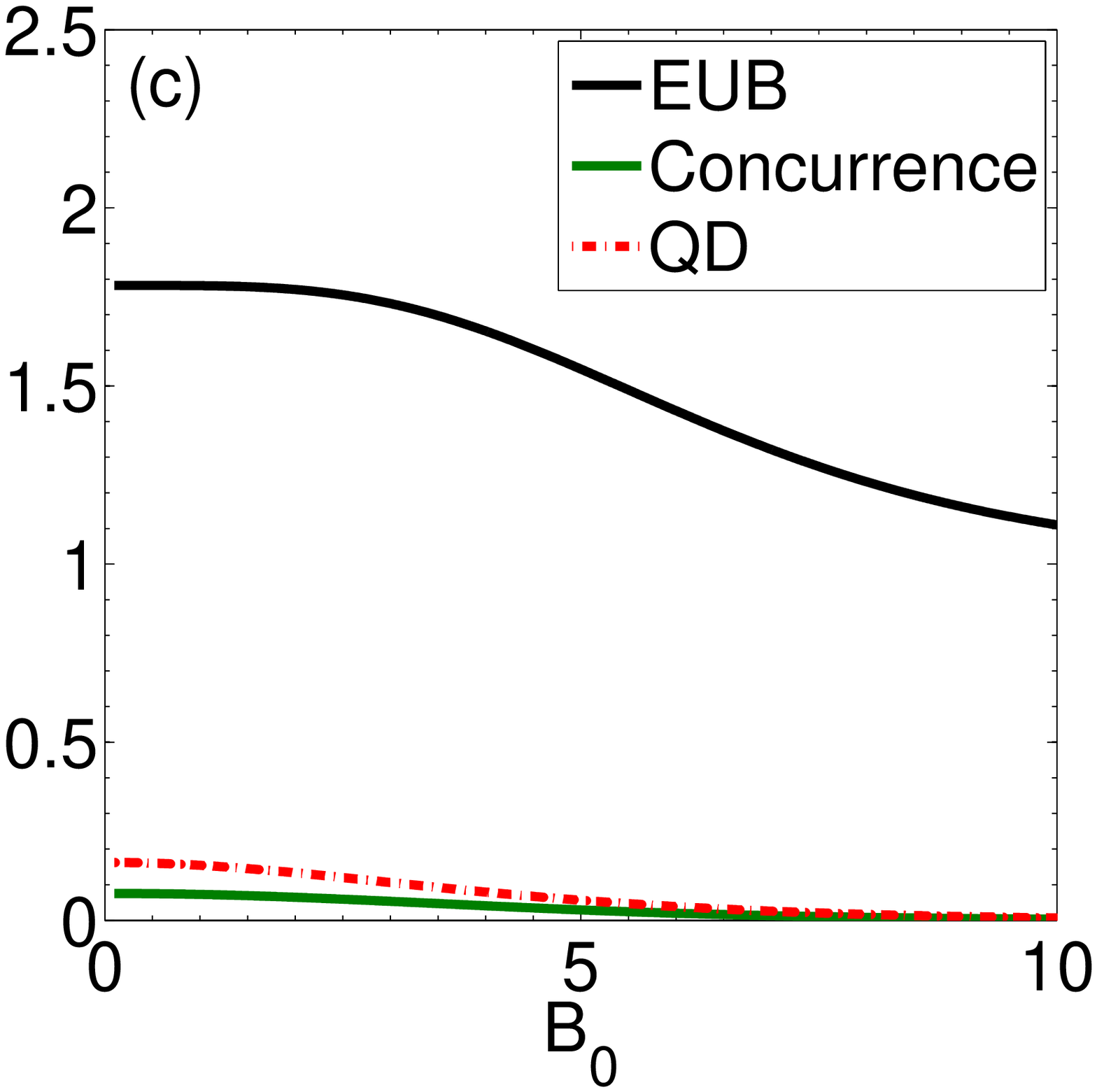}
\caption{Entropic uncertainty bound, concurrence and quantum discord for an isolated quantum
dot in terms of magnetic field $B_0$ with $k_0=10$, $\gamma=1$ (a)$T =0$. (b) $T=1$. (c) $T=2$.}
\label{figure3}
\end{figure}
Now, let us study the EUB, concurrence and quantum discord as a function of the temperature, the deference between the energy of the singlet and triplet states $k_0$, and Magnetic field $B_0$. 

In Fig.\ref{figure1} we have plotted EUB and concurrence and QD versus temperature for specific value of $B_0$ and $k_0$. As can be seen EUB increases with increasing temperature, while quantum discord and concurrence decreases with increasing temperature. Of course, this is quite logical, because with the reduction of quantum correlation between parts $A$ and $B$, the value of uncertainty increases. It is also observed that the EUB increases with decreasing the parameter $k_0$. In Figs. \ref{figure1}(a) and \ref{figure1}(b) we have $\gamma B_0 < k_0/4$, in this situation the ground state at zero temperature is $| \psi_4 \rangle$, which is the maximally entangled state so for $k_0=10$ at zero temperature the system is maximally entangled. As the temperature increases, $| \psi_4 \rangle$ will be mixed with the excited state, so the concurrence and QD monotically decreases from one to zero and uncertainty bound increases from zero to its maximum two. In Fig.\ref{figure1}(c) we have $r>k_0/4$, the ground state of the quantum dot is $| \psi_2 \rangle$, so the concurrence and QD is zero at zero temperature. This state will be mixed with excited state by increasing temperature. So, at first the concurrence and QD increases and then decreases. 

In Fig.\ref{figure2}, we have plotted EUB, concurrence and QD versus parameter $k_0$ for different value of temperature. As can be seen, the EUB start from non-zero value for $k=0$ and reaches to zero for large value of $k_0$. It is observed that at fixed temperature the concurrence and QD increases with increasing the parameter $k_0$. From Figs. \ref{figure2}(a)  \ref{figure2}(b) and  \ref{figure2}(c), it is observed that by increasing the temperature the EUB reaches to zero for larger value of $k_0$. It can also be seen that at fixed temperature when $k_0 =0$ the state of the quantum dot is $| \psi_2 \rangle$, so the concurrence and QD of quantum dot is zero. While for large value of $k_0$ at fixed temperature the  state of the quantum dot is $\vert \psi_4 \rangle$ and the concurrence and QD is equal to one. 

In Fig.\ref{figure3}, we have plotted EUB, concurrence and QD versus magnetic field $B_0$ for different value of temperature. In Fig.\ref{figure3}(a), at zero temperature for $B_0=0$ the state of quantum dot is $| \psi_4 \rangle$ while by increasing $B_0$ the state $| \psi_4 \rangle$ will be mixed with the excited state, so the concurrence and QD monotically decreases from one to zero while EUB increases and reaches to fixed value. In Fig.\ref{figure3}(b), at $T=1$ the state of quantum dot is partially mixed entangled state while for large value of $B_0$ at this temperature the state is $\vert \psi_2 \rangle$ and the entanglement and QD is zero. From Fig. \ref{figure3}(b) it is also observed that the EUB reaches to fixed value for large amount of magnetic field. Finally from Figs. \ref{figure1}(a), \ref{figure1}(b) and \ref{figure1}(c), it is observed that the QD and concurrence decreases with increasing temperature while EUB increases with increasing the temperature. 
\section{conclusion}\label{sec6}
In this work we studied QMA-EUR and quantum correlations in a quantum dot system. We used concurrence  and QD as practical criteria for quantum correlations. We studied the effect of temperature, the energy
difference between the singlet and the triplet states, and the  magnetic field on EUB, concurrence and QD. It was shown that the EUB can be reduced by decreasing the temperature. The situation for quantum coherence is quite different, it grows by reducing the temperature. It is observed that the EUB decreases with increasing the parameter $k_0$, while concurrence and QD increases with increasing this parameter. In studying the effects magnetic field $B_0$ on EUB  it was observed that EUB
grows by increasing the amount of  magnetic field and after touch the summit, they become less to get a steady value for larger value of external magnetic. The concurrence and QD decreases with increasing magnetic field.

\end{document}